\documentclass[11pt]{article}
\usepackage{graphicx}
\usepackage{amsmath}
\usepackage{amsthm}
\usepackage{amsfonts}
\usepackage{algorithm}
\usepackage{adjustbox}
\usepackage{tikz}
\usepackage{bm}
\usepackage{enumitem}
\usepackage{mathtools}

\usepackage{pgfplots} 

\newcommand{\myquote}[1]{``#1''}

\usepackage{array} 

\usepackage{multirow}
\usepackage{booktabs}

\newtheorem{proposition}{Proposition} 
\newtheorem{teo}{Theorem}

{}
\newcommand{\beq}{\begin{equation}}
\newcommand{\eeq}{\end{equation}}
\newcommand{\naturali}{\mathbb{N}}

\newcommand{\ZZ}{\mathbb{Z}}
\newcommand{\RR}{\mathbb{R}}

\newcommand{\opt}[1]{$#1$-OPT}
\newcommand{\kopt}{\opt{k}}
\newcommand{\pref}[1]{(\ref{#1})}
\newcommand{\edg}[2]{\{#1,#2\}}
\newcommand{\PP}[1]{[#1\textrm{?}]}

\newcommand{\Reinss}[3]{<#1,#2,#3>}

\renewcommand{\rq}[1]{r_{#1}}

\newcommand{\orbit}[1]{{\cal O}(#1)}
   
\usepackage{amsmath,amsbsy}

\newcommand{\piu}{\oplus}
\newcommand{\meno}{\ominus}

\newcommand{\falpha}{f_r^1}
\newcommand{\fbeta}{f_r^2}
\newcommand{\fgamma}{f_r^3}
\newcommand{\fdelta}{f_r^4}
\newcommand{\ffalpha}{1}

\begin{document}
	\title{Orbits, schemes and dynamic programming procedures for the TSP \opt{4}\ neighborhood 
		} 		
		\author{Giuseppe Lancia\thanks{DMIF, Univ. of Udine, Italy} \and Marcello Dalpasso\thanks{DEI, Univ. of Padova, Italy}}
\date{}
\maketitle

\begin{abstract}
We discuss the way to group all 25 possible 4-OPT moves into 7 orbits of equivalent moves. We then describe two implementations, one for a $\Theta(n^3)$ algorithm by 
de Berg's et al. and one of a $\Theta(n^2)$ algorithm by
Glover, for finding the best 4-OPT move via dynamic
programming.
\end{abstract}

\newcommand{\bimp}{best-improving}

\section{Introduction}

In this paper we focus on the (symmetric)
Traveling Salesman Problem (TSP), i.e., the problem of finding a shortest hamiltonian cycle (a shortest {\em tour})
on a complete graph of $n$ nodes weighted on the edges.
Let us denote by $c(i,j)=c(j,i)$ the distance between any two nodes $i$ and $j$.  A tour is identified by a permutation of
vertices $(v_1,\ldots,v_n)$. We call $\{v_i, v_{i+1}\}$,  for $i=1,\ldots,n-1$, and $\{v_n,v_1\}$ the {\em edges of the tour}.
The length of a tour $T$, denoted by $c(T)$, is the sum of the lengths of the edges of the tour. More generally, for any set $F$ of edges, we denote by $c(F)$ the value $\sum_{e\in F} c(e)$.

A large number of applications over the years have shown that {\em local search} 
is often a very effective way to tackle hard combinatorial optimization problems \cite{PapSte,AaLe97}.
As far as the TSP is concerned, a famous and successful approach in the past
year has been local search based on the {\em \kopt\ neighborhood}, particularly for $k=2,3$.

Let $k\ge 2$ be an integer constant.
A \kopt\ move on a tour $T$ consists in first removing a set $R$ of $k$ edges and then inserting  a set $I$ of
$k$ edges  so as $(T\setminus R) \cup I$ is still a 
tour.
A \kopt\ move is {\em improving} if
$c((T\setminus R) \cup I) < c(T)$
i.e., $c(I) < c(R)$.
An improving move is {\em \bimp} if $c(R)- c(I)$ is the maximum over all possible choices of $R,I$.
The standard local search approach for the TSP based
on the \kopt\ neighborhood starts from any tour $T^0$ 
(usually a random permutation of the vertices) and then proceeds along a sequence of tours $T^1,T^2,\ldots,T^N$ where each tour $T^j$ is obtained by applying an improving \kopt\ move to $T^{j-1}$. The final tour $T^N$ is such that 
there are no improving \kopt\ moves for it.  The hope is that $T^N$ is a good tour (optimistically, a global optimum) but its quality depends on many factors. One of them is the size of the neighborhood (in this case, $\Theta(n^k)$),
the rationale being that with a larger-size neighborhood we sample a larger number of potential solutions, and hence increase the probability of ending up at a really good one. Clearly, there is a trade-off between the size of a neighborhood and the time required to explore it, so that most times people resort to the use of small neighborhoods since they are very fast to explore.

The first use of \kopt\ dates back to 1958 with the introduction of \opt{2}  in \cite{Croes2opt}.
In 1965  Lin \cite{Lin65} described the  \opt{3} neighborhood, and experimented with a complete enumeration algorithm, of complexity $\Theta(n^3)$, which finds the best \opt{3} move by trying all possibilities. He also introduced a heuristic step fixing some edges of the solution (at risk of being wrong) with the goal of increasing the size of the instances. Still the instances which could be tackled at the time were fairly small ($n\le 150$). Later in 1968, Steiglitz and Weiner \cite{SW} described
an improvement over Lin's method which made it 2 or 3 times faster (although still cubic).

In this paper we focus on the \opt{4} neighborhood, since, in a follow-up work, we are going to describe an effective way for determining the best \opt{4} move, based on a similar approach than the one we followed for exploring the \opt{3} neighborhood \cite{LD20}.

A \opt{4} move starts by removing four edges from a tour, thus creating four separate paths. Then it inserts four edges in such a way as to reconnect the paths into a full tour. There are 25 ways to do so, as we will show later in the paper, so that, basically, there are 25  different types of possible \opt{4} moves. However, many of these moves can be considered equivalent, in the sense that they \myquote{follow the same pattern} (this statement will be made clear later on).
In order to simplify the description of \opt{4} search algorithms, we then formally describe,
in Section \ref{sec:sele}, the equivalence between 
some different moves, and reduce the number of cases to 
consider from 25 to in fact just 7.

In the second part of the paper, we address two algorithms from the literature on the \opt{4} neighborhood. In particular, de Berg et al.
\cite{woeginger} have described a dynamic programming procedure to find the best \opt{4} move in time $\Theta(n^3)$. 
In another paper, Glover \cite{Glover1996} has described a $\Theta(n^2)$ algorithm for finding the best \opt{4} move, but valid only for three particular types of \opt{4} moves. 

The original papers \cite{woeginger,Glover1996} are
not easy reads, and the description of the algorithms can,
in our opinion, be made simpler via the description of dynamic programming relations, which correspond to algorithms of exactly the same complexity as the original versions.
In  Section \ref{sec:literature}, we then describe these dynamic programs, for de Berg et al. and Glover's algorithms.

\section{Selections, schemes and moves}
\label{sec:sele}

Let $G=(V,E)$ be a complete graph on $n$ nodes, and $c:E\mapsto \mathbb{R}^+$ be a cost function for the edges. Without loss of
generality, we assume $V=\{0,1,\ldots,\bar n\}$, where 
$\bar n:=n-1$.  
Furthermore, we always assume the current tour to be the tour $0\rightarrow 1 \rightarrow \cdots \rightarrow \bar n\rightarrow 0$.

We will be using modular arithmetic frequently. For convenience, for each $x\in V$ and $t\in\naturali$ we define
\[
x \piu t := (x + t) \mod n,  \qquad \textrm{} \qquad  x \meno t := (x - t) \mod n.
\]
When moving from $x$ to $x\piu 1, x\piu 2$ etc. we say that we are moving clockwise, or forward. In going from $x$ to $x\meno 1,x\meno 2,\ldots$ we say that we are moving counter-clockwise, or backward.


\newcommand{\selection}{{\cal S}}

A \opt{4}\ move is fully specified by two sets, i.e., the set of
removed and of inserted edges. 
We call a {\em removal set} any set of four tour edges, i.e., four edges of type $\edg{i}{i\piu 1}$. A removal set is identified by a  quadruple $S=(i_1,i_2,i_3,i_4)$ with $0 \le i_1 < i_2  < i_3 < i_4 \le \bar n$, where the edges removed are $R(S):=\{\edg{i_j}{i_j \piu 1} : j=1,\ldots,4\}$. We call any such quadruple $S$ a {\em selection}.
A selection is {\em complete} if
$i_{h}\piu 1\notin\{i_1,\ldots,i_4\}$ for each $h=1,\ldots,4$ (i.e., if the move never removes two consecutive edges of the tour), otherwise we say that $S$ is a {\em partial} selection. 
Complete selections should be distinguished from partial \opt{4}  selections, since the number of choices required to determine a partial selection is actually lower than 
four. For instance, there is only a cubic number of selections in which $i_4=i_3\piu 1$ since we can choose $i_1$, $i_2$ and $i_3$ but the value of $i_4$ is forced.

\newcommand{\reinsset}{{\cal I}}

Let $S$ be a selection and $I\subset E$ with $|I|=4$. If $(T \setminus R(S)) \cup I$ is still a tour then $I$ is called a {\em  reinsertion set}. 
Given a selection $S$, a reinsertion set $I$ is {\em pure} if $I\cap R(S) = \emptyset$, and {\em degenerate} otherwise. 
Finding the best \opt{4}\ move when the reinsertions are constrained to be degenerate is $O(n^3)$ (in fact, \opt{4} degenerates to either \opt{2} or \opt{3} in this case). Therefore, the most computationally expensive task is to determine the best move when {\em the selection is complete and the reinsertion is pure}.
We  refer to this kind of moves as {\em true} \opt{4}. 
Thus, in the remainder of the paper we will focus on true  \opt{4}\ moves.

 \subsection{Symmetries and orbits of reinsertion sets}

Let $\selection$ be the set of all complete selections. 
For $S\in \selection$, let us denote by $\reinsset(S)$ the set of all
pure reinsertion sets for $S$. 
Then, the set of all true \opt{4}\ moves is
\beq
\big\{\big(R(S),I\big) : S\in \selection, I\in \reinsset(S)\big\}
\label{set:allmoves}
\eeq
and the total number of true \opt{4} moves is
$\sum_{S\in\selection} |\reinsset(S)| = \gamma \,|\selection|$,
where  $\gamma$
specifies in how many different ways a tour can be reassembled back after four edges have been removed and replaced by four different ones.
We will see in section \ref{sec:comb} that this number, a constant over all the selections $S$, equals $25$ for \opt{4}.

According to \pref{set:allmoves}, to list all true \opt{4}\ moves we could consider a case analysis in which we go through all pure reinsertion sets for all complete selections. We can, however, simplify this case analysis by grouping together different reinsertion sets which are, in fact, equivalent. 

To define the equivalence between reinsertion sets, let us introduce a standard way to draw a tour with a reinsertion set. This type of figures will be in 1-to-1 correspondence with the reinsertion sets.  We  will then exploit the symmetry between the drawings in order to partition the  reinsertion sets into equivalence classes.

\begin{figure}[t]
\begin{center}
	\begin{tikzpicture}[scale=0.55]
		\draw[ultra thick] (-0.351,1.765) arc (101.25:168.75:1.8);
		\filldraw (1.765,0.351) circle (2pt);
		\filldraw (1.765,-0.351) circle (2pt);
		\node at (-0.456,2.295) {$1$};
		\node at (0.456,2.295) {$1'$};
		\draw[ultra thick] (1.765,0.351) arc (11.25:78.75:1.8);
		\filldraw (0.351,-1.765) circle (2pt);
		\filldraw (-0.351,-1.765) circle (2pt);
		\node at (2.295,0.456) {$2$};
		\node at (2.295,-0.456) {$2'$};
		\draw[ultra thick] (0.351,-1.765) arc (-78.75:-11.25:1.8);
		\filldraw (-1.765,-0.351) circle (2pt);
		\filldraw (-1.765,0.351) circle (2pt);
		\node at (0.456,-2.295) {$3$};
		\node at (-0.456,-2.295) {$3'$};
		\draw[ultra thick] (-1.765,-0.351) arc (-168.75:-101.25:1.8);
		\filldraw (-0.351,1.765) circle (2pt);
		\filldraw (0.351,1.765) circle (2pt);
		\node at (-2.295,-0.456) {$4$};
		\node at (-2.295,0.456) {$4'$};
		
	\tikzset{shift={(-4,0)}}
		
		\draw[ultra thick] (11.648,1.765) arc (101.25:168.75:1.8);
		\filldraw (13.765,0.351) circle (2pt);
		\filldraw (13.765,-0.351) circle (2pt);
		\node at (11.543,2.295) {$1$};
		\node at (12.456,2.295) {$1'$};
		\draw[ultra thick] (13.765,0.351) arc (11.25:78.75:1.8);
		\filldraw (12.351,-1.765) circle (2pt);
		\filldraw (11.648,-1.765) circle (2pt);
		\node at (14.295,0.456) {$2$};
		\node at (14.295,-0.456) {$2'$};
		\draw[ultra thick] (12.351,-1.765) arc (-78.75:-11.25:1.8);
		\filldraw (10.234,-0.351) circle (2pt);
		\filldraw (10.234,0.351) circle (2pt);
		\node at (12.456,-2.295) {$3$};
		\node at (11.543,-2.295) {$3'$};
		\draw[ultra thick] (10.234,-0.351) arc (-168.75:-101.25:1.8);
		\filldraw (11.648,1.765) circle (2pt);
		\filldraw (12.351,1.765) circle (2pt);
		\node at (9.704,-0.456) {$4$};
		\node at (9.704,0.456) {$4'$};
		\draw [thick] (11.648,1.765) -- (13.765,0.351);
		\draw [thick] (12.351,1.765) -- (11.648,-1.765);
		\draw [thick] (10.234,-0.351) -- (12.351,-1.765);
		\draw [thick] (13.765,-0.351) -- (10.234,0.351);

	\end{tikzpicture}
\end{center}
\label{fig:draw4}
\caption{Graphical representation of a \opt{4} move}
\end{figure}
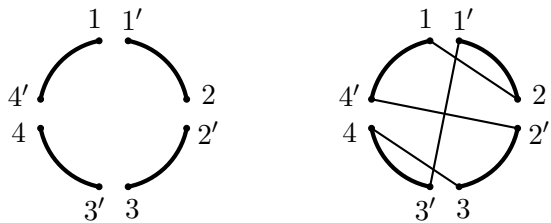

Take two angles $\alpha$ and $\beta$ such that $\alpha+\beta=\pi/2$ (for a nice drawing, $\alpha$ should be quite smaller than $\beta$, e.g., $\alpha= \beta/4$). Along an invisible circle, starting with an arc centered at $(\pi-\alpha)/2$, alternate the following: skip an arc of angle $\alpha$ and draw an arc of angle $\beta$. This will produce $4$ arcs equally spaced along the circle.  The endpoints of these arcs are labeled, starting at the top and proceeding clockwise, as $1,1',2,2',\ldots, 4,4'$ (see Fig. \ref{fig:draw4}, left).

Assuming that a full circle represents the tour before the move, 
the points labeled $1,1',\ldots, 4,4'$ are associated to the selection indices
of a generic complete selection $S=(i_1,\ldots, i_4)$ (where a label $l$ represents node $i_l$  
while a label $l'$ represents node $i_l\piu 1$). The  
 arcs $\edg{l}{l'}$ missing from the drawing represent the edges that were removed.

A reinsertion set corresponds to four lines inside the circle which match the points in $\{1,1',\ldots,4,4'\}$ and complete the drawing of a tour (see Fig. \ref{fig:draw4}, right).
Our task will be to identify all such drawings. To simplify the exposition and reduce the number of cases to consider we can exploit the fact that some drawings are symmetric and can be put in one equivalence class.

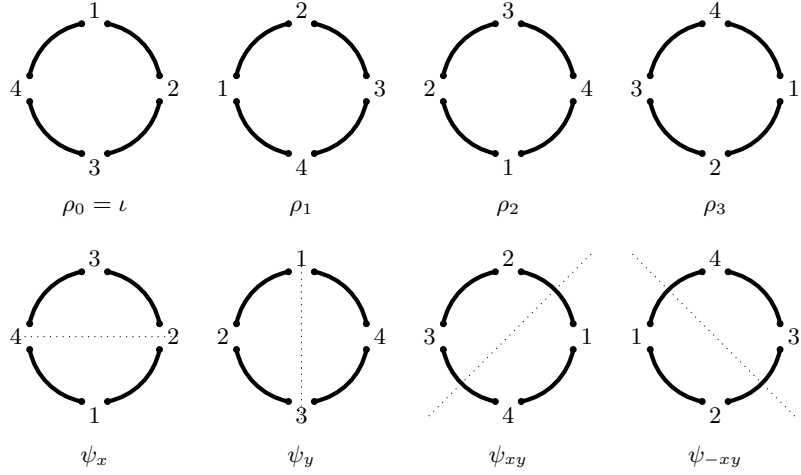
\begin{figure}[h]
\begin{center}
	{\footnotesize
		\begin{tikzpicture}[scale=0.55]
			
			\draw[ultra thick] (1.569,0.312) arc (11.25:78.75:1.6);
			\filldraw (0.312,-1.569) circle (2pt);
			\filldraw (-0.312,-1.569) circle (2pt);
			\draw[ultra thick] (0.312,-1.569) arc (-78.75:-11.25:1.6);
			\filldraw (-1.569,-0.312) circle (2pt);
			\filldraw (-1.569,0.312) circle (2pt);
			\draw[ultra thick] (-1.569,-0.312) arc (-168.75:-101.25:1.6);
			\filldraw (-0.312,1.569) circle (2pt);
			\filldraw (0.312,1.569) circle (2pt);
			\draw[ultra thick] (-0.312,1.569) arc (-258.75:-191.25:1.6);
			\filldraw (1.569,0.312) circle (2pt);
			\filldraw (1.569,-0.312) circle (2pt);

			\node at (0,1.9) {$3$};
			\node at (5,1.9) {$1$};
			\node at (10,1.9) {$2$};
			\node at (15,1.9) {$4$};
			
			\node at (0, -1.9) {$1$}; 
			\node at (5, -1.9) {$3$}; 
			\node at (10, -1.9) {$4$}; 
			\node at (15, -1.9) {$2$}; 
			
			\node at (1.9,0) {2};
			\node at (6.9,0) {4}; 
			\node at (11.9,0) {1}; 
			\node at (16.9,0) {3}; 
			
			\node at (-1.9,0) {4};
			\node at (3.1,0) {2}; 
			\node at (8.1,0) {3}; 
			\node at (13.1,0) {1};

			\node at (0,7.9) {$1$};
			\node at (5,7.9) {$2$};
			\node at (10,7.9) {$3$};
			\node at (15,7.9) {$4$};
			
			\node at (0, 4.1) {$3$}; 
			\node at (5, 4.1) {$4$}; 
			\node at (10, 4.1) {$1$}; 
			\node at (15, 4.1) {$2$}; 
			
			\node at (1.9,6) {2};
			\node at (6.9,6) {3}; 
			\node at (11.9,6) {4}; 
			\node at (16.9,6) {1}; 
			
			\node at (-1.9,6) {4};
			\node at (3.1,6) {1}; 
			\node at (8.1,6) {2}; 
			\node at (13.1,6) {3}; 
			
			\node at (0, 3.1) {$\rho_0=\iota$}; 
			\node at (5, 3.1) {$\rho_1$}; 
			\node at (10, 3.1) {$\rho_2$};      
			\node at (15, 3.1) {$\rho_3$}; 
			\node at (0, -2.9) {$\psi_x$}; 
			\node at (5, -2.9) {$\psi_y$}; 
			\node at (10, -2.9) {$\psi_{xy}$};      
			\node at (15, -2.9) {$\psi_{-xy}$}; 
			
			\draw[ultra thick] (6.569,0.312) arc (11.25:78.75:1.6);
			\filldraw (5.312,-1.569) circle (2pt);
			\filldraw (4.687,-1.569) circle (2pt);
			\draw[ultra thick] (5.312,-1.569) arc (-78.75:-11.25:1.6);
			\filldraw (3.43,-0.312) circle (2pt);
			\filldraw (3.43,0.312) circle (2pt);
			\draw[ultra thick] (3.43,-0.312) arc (-168.75:-101.25:1.6);
			\filldraw (4.687,1.569) circle (2pt);
			\filldraw (5.312,1.569) circle (2pt);
			\draw[ultra thick] (4.687,1.569) arc (-258.75:-191.25:1.6);
			\filldraw (6.569,0.312) circle (2pt);
			\filldraw (6.569,-0.312) circle (2pt);

			\draw[ultra thick] (11.569,0.312) arc (11.25:78.75:1.6);
			\filldraw (10.312,-1.569) circle (2pt);
			\filldraw (9.687,-1.569) circle (2pt);
			\draw[ultra thick] (10.312,-1.569) arc (-78.75:-11.25:1.6);
			\filldraw (8.43,-0.312) circle (2pt);
			\filldraw (8.43,0.312) circle (2pt);
			\draw[ultra thick] (8.43,-0.312) arc (-168.75:-101.25:1.6);
			\filldraw (9.687,1.569) circle (2pt);
			\filldraw (10.312,1.569) circle (2pt);
			\draw[ultra thick] (9.687,1.569) arc (-258.75:-191.25:1.6);
			\filldraw (11.569,0.312) circle (2pt);
			\filldraw (11.569,-0.312) circle (2pt);
			
			\draw[ultra thick] (16.569,0.312) arc (11.25:78.75:1.6);
			\filldraw (15.312,-1.569) circle (2pt);
			\filldraw (14.687,-1.569) circle (2pt);
			\draw[ultra thick] (15.312,-1.569) arc (-78.75:-11.25:1.6);
			\filldraw (13.43,-0.312) circle (2pt);
			\filldraw (13.43,0.312) circle (2pt);
			\draw[ultra thick] (13.43,-0.312) arc (-168.75:-101.25:1.6);
			\filldraw (14.687,1.569) circle (2pt);
			\filldraw (15.312,1.569) circle (2pt);
			\draw[ultra thick] (14.687,1.569) arc (-258.75:-191.25:1.6);
			\filldraw (16.569,0.312) circle (2pt);
			\filldraw (16.569,-0.312) circle (2pt);

			\draw[ultra thick] (1.569,6.312) arc (11.25:78.75:1.6);
			\filldraw (0.312,4.43) circle (2pt);
			\filldraw (-0.312,4.43) circle (2pt);
			\draw[ultra thick] (0.312,4.43) arc (-78.75:-11.25:1.6);
			\filldraw (-1.569,5.687) circle (2pt);
			\filldraw (-1.569,6.312) circle (2pt);
			\draw[ultra thick] (-1.569,5.687) arc (-168.75:-101.25:1.6);
			\filldraw (-0.312,7.569) circle (2pt);
			\filldraw (0.312,7.569) circle (2pt);
			\draw[ultra thick] (-0.312,7.569) arc (-258.75:-191.25:1.6);
			\filldraw (1.569,6.312) circle (2pt);
			\filldraw (1.569,5.687) circle (2pt);

			\draw[ultra thick] (6.569,6.312) arc (11.25:78.75:1.6);
			\filldraw (5.312,4.43) circle (2pt);
			\filldraw (4.687,4.43) circle (2pt);
			\draw[ultra thick] (5.312,4.43) arc (-78.75:-11.25:1.6);
			\filldraw (3.43,5.687) circle (2pt);
			\filldraw (3.43,6.312) circle (2pt);
			\draw[ultra thick] (3.43,5.687) arc (-168.75:-101.25:1.6);
			\filldraw (4.687,7.569) circle (2pt);
			\filldraw (5.312,7.569) circle (2pt);
			\draw[ultra thick] (4.687,7.569) arc (-258.75:-191.25:1.6);
			\filldraw (6.569,6.312) circle (2pt);
			\filldraw (6.569,5.687) circle (2pt);

			\draw[ultra thick] (11.569,6.312) arc (11.25:78.75:1.6);
			\filldraw (10.312,4.43) circle (2pt);
			\filldraw (9.687,4.43) circle (2pt);
			\draw[ultra thick] (10.312,4.43) arc (-78.75:-11.25:1.6);
			\filldraw (8.43,5.687) circle (2pt);
			\filldraw (8.43,6.312) circle (2pt);
			\draw[ultra thick] (8.43,5.687) arc (-168.75:-101.25:1.6);
			\filldraw (9.687,7.569) circle (2pt);
			\filldraw (10.312,7.569) circle (2pt);
			\draw[ultra thick] (9.687,7.569) arc (-258.75:-191.25:1.6);
			\filldraw (11.569,6.312) circle (2pt);
			\filldraw (11.569,5.687) circle (2pt);

			\draw[ultra thick] (16.569,6.312) arc (11.25:78.75:1.6);
			\filldraw (15.312,4.43) circle (2pt);
			\filldraw (14.687,4.43) circle (2pt);
			\draw[ultra thick] (15.312,4.43) arc (-78.75:-11.25:1.6);
			\filldraw (13.43,5.687) circle (2pt);
			\filldraw (13.43,6.312) circle (2pt);
			\draw[ultra thick] (13.43,5.687) arc (-168.75:-101.25:1.6);
			\filldraw (14.687,7.569) circle (2pt);
			\filldraw (15.312,7.569) circle (2pt);
			\draw[ultra thick] (14.687,7.569) arc (-258.75:-191.25:1.6);
			\filldraw (16.569,6.312) circle (2pt);
			\filldraw (16.569,5.687) circle (2pt);
			

			\draw [dotted] (5,-2) -- (5,2);
			\draw [dotted] (-2,0) -- (2,0);
			\draw [dotted] (13,2) -- (17,-2);
			\draw [dotted] (12,2) -- (8,-2);
			
		\end{tikzpicture}
	}
\caption{The set of all symmetry operators applied to a tour}
	\label{fig:symmetries}
\end{center}
\end{figure}

In general, symmetries of figures in the plane are defined by rotations and reflection along axes of symmetry. In our case there are four rotations and four reflections to consider. 
For $j=0,\ldots,3$ let us denote by $\rho_j$ the  rotation of the labels by $j\pi/2$ counterclockwise. Furthermore, we denote by $\psi_x$, $\psi_y$, $\psi_{xy}$ and $\psi_{-xy}$ the reflections along, respectively, the $x$-axis, the $y$-axis, the $(x=y)$-line and the $(x=-y)$-line. The permutations corresponding to these operators are depicted, graphically, in Fig. \ref{fig:symmetries}.

The set ${\cal G}=\{\rho_0,\rho_1,\rho_2,\rho_3,\psi_{x} 
,\psi_{y},\psi_{xy},\psi_{-xy}\}$ is a group, of order four, under the operation of function composition. In algebra it is known as the {\em octic} group. The identity is $\rho_0$. Let $\rho=\rho_1$ and $\psi=\psi_x$. Then we have $\rho_j=\rho^j$ and $\psi_y = \psi \rho^2$, 
$\psi_{xy} = \psi \rho^3$, 
$\psi_{-xy} = \psi \rho$, so that 
${\cal G}=\{\rho^j,\psi\rho^j, j=0,\ldots,3\}$.
To define equivalent (i.e., symmetric) reinsertion sets,
we  need to apply some results from the algebraic theory of group actions. We define a group action of ${\cal G}$ on $\reinsset(S)$ by specifying, for each $\varphi\in {\cal G}$ and $I\in \reinsset(S)$ a set $\varphi I\in\reinsset(S)$. For each $\varphi\in {\cal G}$, this association must be  a bijection of  $\reinsset(S)$ into itself.

Let us then define new labels for each node $x$, $x'$ with $x=1,\ldots,4$, under the action of $\varphi$ as
\beq
\begin{cases}
	L_{\varphi}(x)=\varphi(x)\quad\ L_{\varphi}(x') =\varphi(x)' \qquad \textrm{ if } \varphi=\rho^j \textrm{ for some } j\\
	L_{\varphi}(x)=\varphi(x)'\quad L_{\varphi}(x') =\varphi(x)\, \qquad \textrm{ otherwise}\\
\end{cases}
\label{def:L}
\eeq
Then a reinsertion set $I$ is mapped by $\varphi$ into the reinsertion set
\[
\varphi I :=\big\{\edg{L_{\varphi}(x)}{L_{\varphi}(y)} : \edg{x}{y}\in I\big\}
\]

Two reinsertion sets $I$ and $I'$ are said to be symmetric if there exists $\varphi\in{\cal G}$ such that $I'=\varphi I$. This relation is in fact an equivalence. The equivalence class $\orbit{I}=\{\varphi I:  \varphi\in {\cal G}\}$ is called the {\em orbit} of $I$. The orbits partition the set $\reinsset(S)$. It is a known result from algebra that in a group action the size of an orbit must divide the size of the group. Therefore, in \opt{4} each orbit  must have 1, 2, 4 or 8 elements.

\begin{figure}[t]
\begin{center}
	\begin{tikzpicture}[scale=0.55]
		\draw[ultra thick] (-0.39,1.961) arc (101.25:168.75:2);
		\filldraw (1.961,0.39) circle (2pt);
		\filldraw (1.961,-0.39) circle (2pt);
		\draw [dotted,thick] (-0.39,1.961) -- (0.39,1.961);
		\node at (-0.507,2.55) {{\small $1$}};
		\node at (0.507,2.55) {\small $1'$};
		\draw[ultra thick] (1.961,0.39) arc (11.25:78.75:2);
		\filldraw (0.39,-1.961) circle (2pt);
		\filldraw (-0.39,-1.961) circle (2pt);
		\draw [dotted,thick] (1.961,0.39) -- (1.961,-0.39);
		\node at (2.55,0.507) {\small $2$};
		\node at (2.55,-0.507) {\small $2'$};
		\draw[ultra thick] (0.39,-1.961) arc (-78.75:-11.25:2);
		\filldraw (-1.961,-0.39) circle (2pt);
		\filldraw (-1.961,0.39) circle (2pt);
		\draw [dotted,thick] (0.39,-1.961) -- (-0.39,-1.961);
		\node at (0.507,-2.55) {\small $3$};
		\node at (-0.507,-2.55) {\small $3'$};
		\draw[ultra thick] (-1.961,-0.39) arc (-168.75:-101.25:2);
		\filldraw (-0.39,1.961) circle (2pt);
		\filldraw (0.39,1.961) circle (2pt);
		\draw [dotted,thick] (-1.961,-0.39) -- (-1.961,0.39);
		\node at (-2.55,-0.507) {\small $4$};
		\node at (-2.55,0.507) {\small $4'$};
		\draw [thick] (-0.39,1.961) -- (-0.39,-1.961);
		\draw [thick] (-1.961,-0.39) -- (1.961,0.39);
		\draw [thick] (0.39,1.961) -- (0.39,-1.961);
		\draw [thick] (1.961,-0.39) -- (-1.961,0.39);
		
		\draw[ultra thick] (6.609,1.961) arc (101.25:168.75:2);
		\filldraw (8.961,0.39) circle (2pt);
		\filldraw (8.961,-0.39) circle (2pt);
		\draw [dotted,thick] (6.609,1.961) -- (7.39,1.961);
		\node at (6.492,2.55) {\small $1$};
		\node at (7.507,2.55) {\small $1'$};
		\draw[ultra thick] (8.961,0.39) arc (11.25:78.75:2);
		\filldraw (7.39,-1.961) circle (2pt);
		\filldraw (6.609,-1.961) circle (2pt);
		\draw [dotted,thick] (8.961,0.39) -- (8.961,-0.39);
		\node at (9.55,0.507) {\small $2$};
		\node at (9.55,-0.507) {\small $2'$};
		\draw[ultra thick] (7.39,-1.961) arc (-78.75:-11.25:2);
		\filldraw (5.038,-0.39) circle (2pt);
		\filldraw (5.038,0.39) circle (2pt);
		\draw [dotted,thick] (7.39,-1.961) -- (6.609,-1.961);
		\node at (7.507,-2.55) {\small $3$};
		\node at (6.492,-2.55) {\small $3'$};
		\draw[ultra thick] (5.038,-0.39) arc (-168.75:-101.25:2);
		\filldraw (6.609,1.961) circle (2pt);
		\filldraw (7.39,1.961) circle (2pt);
		\draw [dotted,thick] (5.038,-0.39) -- (5.038,0.39);
		\node at (4.449,-0.507) {\small $4$};
		\node at (4.449,0.507) {\small $4'$};
		\draw [thick] (6.609,1.961) -- (7.39,-1.961);
		\draw [thick] (8.961,-0.39) -- (5.038,-0.39);
		\draw [thick] (6.609,-1.961) -- (7.39,1.961);
		\draw [thick] (8.961,0.39) -- (5.038,0.39);
		
	\end{tikzpicture}
\end{center}
\caption{A reinsertion set before and after a rotation}
\label{fig:rot}
\end{figure}
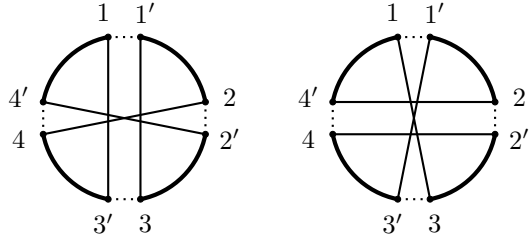

Consider, for example, the reinsertion set $\{\edg{i_1}{i_3\piu 1}, \edg{i_1\piu 1}{i_3},\edg{i_2}{i_4}, \edg{i_2\piu 1}{i_4\piu 1}\}$ which, in our notation, is
\[
I=\{\edg{1}{3'}, \edg{1'}{3},\edg{2}{4}, \edg{2'}{4'}\}
\]
and is depicted in Fig. \ref{fig:rot}(left). 
If we apply a rotation $\rho$ we obtain the set
\begin{align*}
\rho I &=\{\edg{\rho({1})}{\rho({3})'}, \edg{\rho({1})'}{\rho({3})},\edg{\rho({2})}{\rho({4})}, \edg{\rho({2})'}{\rho({4})'}\}\\
 & =\{\edg{2}{4'}, \edg{2'}{4},\edg{3}{1}, \edg{3'}{1'}\}
\end{align*}

which is depicted in Fig. \ref{fig:rot}(right) and corresponds to the actual edges 
$\{\edg{i_1}{i_3}, \edg{i_1\piu 1}{i_3\piu 1},\edg{i_2}{i_4\piu 1}, \edg{i_2\piu 1}{i_4}\}$. If we apply the rotation $\rho$ again we obtain 
\begin{align*}
\rho^2 I & = \{\edg{\rho({2})}{\rho(4)'}, \edg{\rho(2)'}{\rho({4})},\edg{\rho({3})}{\rho({1})}, \edg{\rho({3})'}{\rho({1})'}\}\\
& =\{\edg{3}{1'}, \edg{3'}{1},\edg{4}{2}, \edg{4'}{2'}\}\\
& =I
\end{align*}
Similarly,
\begin{align*}
\psi I &=\{\edg{\psi(1)'}{\psi({3})}, \edg{\psi(1)}{\psi({3})'},\edg{\psi({2})'}{\psi({4})'}, \edg{\psi({2})}{\psi({4})}\}\\
 & =\{\edg{3'}{1}, \edg{3}{1'},\edg{2}{4}, \edg{2'}{4'}\}\\
 & =I
\end{align*}

Therefore $\orbit{I}=\{I,\rho I\}$ and the orbit has size 2.

In order  to enumerate all the reinsertion sets so as to compute their orbits and partition them into a small number of cases, we introduce a handy notation called {\em reinsertion schemes}. Reinsertion schemes will be in 1-to-1 correspondence with reinsertion sets.

\subsection{Reinsertion schemes}
\label{sec:comb}

Let $S$ be a complete \opt{4} selection.
When the edges $R(S)$ are removed from a tour, the tour
gets broken into four segments  which we label by $\{1,\ldots,4\}$. For $l=1,\ldots,4$, the  segment labeled $l$ is the path that has  $i_l$ as its last vertex. In particular, the segments are $(i_4\piu 1,\ldots, i_1)$, $(i_1\piu 1,\ldots,  i_2)$,
$(i_2\piu 1,\ldots, i_3)$ and $(i_3\piu 1,\ldots, i_4)$.
Since the selection is pure, each segment contains at least one edge. A reinsertion set 
patches back these segments into a new tour. If we adopt the convention to start always a tour with segment 1 traversed clockwise, 
the reinsertion set: (i) determines a new ordering in which the segments are visited along the tour and (ii) may cause some segments to be traversed counterclockwise.
In order to represent this fact, instead of listing the edges of a reinsertion set we can use an alternative notation called a {\em reinsertion scheme}. A reinsertion scheme is a signed permutation of $\{2,3,4\}$. The permutation specifies the order in which the segments $2,3,4$ are visited after the move. The signing $-s$ tells that segment $s$ is traversed counterclockwise, while $+s$ tells that it is traversed clockwise. 
For example, the reinsertion set depicted in Fig. \ref{fig:rot}(left) is also represented by the reinsertion scheme
$\Reinss{+4}{-2}{-3}$ since  from the end of segment 1  we jump to the beginning of segment 4 and traverse the segment forward. We then move to the last element of  segment 2 and proceed backward to its first element. We then jump to the end of segment 3 and proceed backward to its beginning. Finally, we close the cycle by going back to the first element of segment 1.

Clearly, there is a bijection between reinsertion schemes and reinsertion sets. If $r$ is a reinsertion scheme,  we denote by $I(r)$ 
the corresponding reinsertion set, while if $I$ is a reinsertion set, we denote by $r(I)$ the corresponding reinsertion scheme. We can then readily extend the idea of orbits and group actions to reinsertion schemes. Namely, we will consider the orbit of a reinsertion scheme to be the set of all reinsertion schemes whose corresponding reinsertion sets are symmetric. Let $I$ be a reinsertion set, and $r=r(I)$. Then  we extend the action of ${\cal G}$ to reinsertion schemes by defining, for each $\varphi\in{\cal G}$, the scheme $\varphi r$ to be
$r(\varphi I)$. For example, by looking at Fig. \ref{fig:rot}(b) we have that $\rho \Reinss{-2}{+4}{-3} = \Reinss{-3}{-4}{+2}$.
Because of the equivalence between reinsertion sets and reinsertion schemes, in the following we will be using either of them, at our convenience, for the sake of simplicity. 

There are potentially $2^{3}\times 3!$ reinsertion schemes for \opt{4}, but for many of these the corresponding reinsertion sets are degenerate. A scheme for a pure reinsertion  must not  start with $+2$, nor end with ``$+4$'', nor contain consecutive elements ``$+t,+(t+1)$'' or ``$-t,-(t-1)$'' for any $t$ in $1,\ldots,4$.

\begin{proposition}
	There are 25 pure reinsertion schemes for \opt{4}.
\end{proposition}

\begin{proof} 
The pure schemes, classified by a permutation $\pi$ of $\{2,3,4\}$ first and then by the signing, are the following:
	\begin{enumerate}
		\item [-] $\pi=(2,3,4)$: Signing $+2$ is forbidden, and also $+4$ is forbidden. This leaves only two possibilities
		\begin{enumerate}
			\item [] $\rq{1}=\Reinss{-2}{-3}{-4}$\qquad \qquad $\rq{2}=\Reinss{-2}{+3}{-4}$
		\end{enumerate}
		\item [-] $\pi=(2,4,3)$: Signing $+2$ is forbidden. Also the sequence $-4,-3$ is forbidden. This leaves three possibilities
		\begin{enumerate}
			\item [] $\rq{3}=\Reinss{-2}{-4}{+3}$ \qquad \qquad
			$\rq{4} = \Reinss{-2}{+4}{-3}$ \qquad \qquad
			$\rq{5} = \Reinss{-2}{+4}{+3}$
		\end{enumerate}
		
		\item [-] $\pi=(3,2,4)$: Signing $+4$ is forbidden. Also the sequence $-3,-2$ is forbidden. This leaves three possibilities
		\begin{enumerate}
			\item [] $\rq{6} = \Reinss{-3}{+2}{-4}$ \qquad \qquad
			$\rq{7} = \Reinss{+3}{-2}{-4}$ \qquad \qquad
			$\rq{8} = \Reinss{+3}{+2}{-4}$
		\end{enumerate}
		
		\item [-] $\pi=(3,4,2)$: The sequence $+3,+4$ is forbidden. This leaves six possibilities
		\begin{enumerate}
			\item[] $\rq{9} = \Reinss{-3}{-4}{-2}$ \qquad \qquad
			$\rq{10} = \Reinss{-3}{-4}{+2}$ \qquad \qquad
			$\rq{11} = \Reinss{-3}{+4}{-2}$ 
			\item [] $\rq{12} = \Reinss{-3}{+4}{+2}$ \qquad \quad\
			$\rq{13} = \Reinss{+3}{-4}{-2}$ \qquad \qquad
			$\rq{14} = \Reinss{+3}{-4}{+2}$
		\end{enumerate}
		%
		%
		\item [-] $\pi=(4,2,3)$: The sequence $+2,+3$ is forbidden. This leaves six possibilities
		\begin{enumerate}
			\item[] $\rq{15} = \Reinss{-4}{-2}{-3}$ \qquad \qquad
			$\rq{16} = \Reinss{+4}{-2}{-3}$ \qquad \qquad
			$\rq{17} = \Reinss{-4}{-2}{+3}$
			\item[] $\rq{18} = \Reinss{+4}{-2}{+3}$ \qquad \qquad
			$\rq{19} = \Reinss{-4}{+2}{-3}$\qquad \qquad
			$\rq{20} = \Reinss{+4}{+2}{-3}$
		\end{enumerate}
		%
		%
		\item [-] $\pi=(4,3,2)$ : The sequence $-4,-3$ is forbidden as well as the sequence $-3,-2$. This leaves five possibilities
		\begin{enumerate}
			\item[] $\rq{21} = \Reinss{-4}{+3}{-2}$ \qquad \qquad
			$\rq{22} = \Reinss{-4}{+3}{+2}$ \qquad \qquad
			$\rq{23} = \Reinss{+4}{-3}{+2}$
			\item[] $\rq{24} = \Reinss{+4}{+3}{-2}$ \qquad \qquad
			$\rq{25} = \Reinss{+4}{+3}{+2}$
		\end{enumerate}
		%
		
	\end{enumerate}
\end{proof}
We can now proceed and compute the orbits for the pure reinsertion schemes.

\begin{proposition}
	The pure reinsertion schemes for \opt{4} are partitioned in  7 orbits ${\cal O}_1,\ldots,{\cal O}_7$.
\end{proposition}
\begin{proof}
	We have 
	\begin{enumerate}
		\item [-] ${\cal O}_1=\orbit{\rq{1}}=\{\rq{1},\rq{24},\rq{23},\rq{22}\} = 
		\{\rq{1}, \rho\rq{1}, \rho^2\rq{1}, \rho^3\rq{1}\}$.
		\item [-] ${\cal O}_2=\orbit{\rq{2}}=\{\rq{2},\rq{21}\} = \{\rq{2},\rho\rq{2}\}$.
		\item [-] ${\cal O}_3=\orbit{\rq{3}}=\{\rq{3},\rq{7},\rq{13},\rq{17}\} = \{\rq{3},\rho\rq{3},\rho^2\rq{3}, \rho^3\rq{3}\}$.
		\item [-] ${\cal O}_4=\orbit{\rq{4}}=\{\rq{4},\rq{19},\rq{11},\rq{6}\} = \{\rq{4},\rho\rq{4},\psi\rq{4}, \psi\rho\rq{4}\}$.
		\item [-] ${\cal O}_5=\orbit{\rq{5}}=\{\rq{5},\rq{20},\rq{14},\rq{15},\rq{12},\rq{18},\rq{9},\rq{8}\} = \{\rq{5},\rho\rq{5},\rho^2\rq{5},\rho^3\rq{5},\psi\rq{5}, \psi\rho\rq{5},\psi\rho^2\rq{5}, \psi\rho^3\rq{5}\}$.
		\item [-] ${\cal O}_6=\orbit{\rq{10}}=\{\rq{10},\rq{16}\}=\{\rq{10},\rho\rq{10}\}$.
		\item [-] ${\cal O}_7=\orbit{\rq{25}}=\{\rq{25}\}$.
	\end{enumerate}
\end{proof}

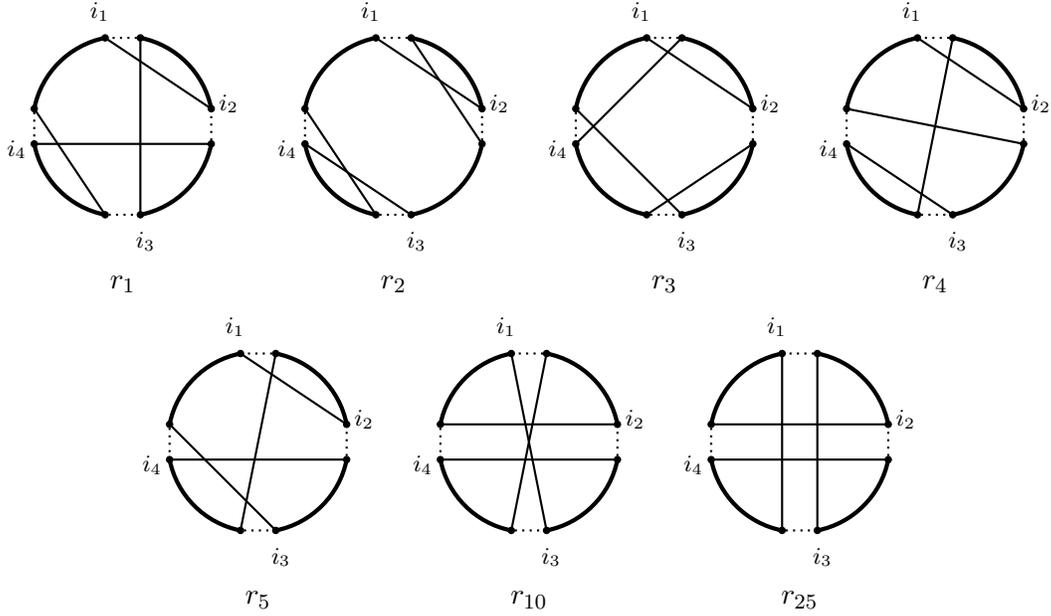
\begin{figure}[tb]
	\begin{center}
		{\footnotesize
			\begin{tikzpicture}[scale=0.6]
				\draw[ultra thick] (-0.39,1.961) arc (101.25:168.75:2);
				\filldraw (1.961,0.39) circle (2pt);
				\filldraw (1.961,-0.39) circle (2pt);
				\draw [dotted,thick] (-0.39,1.961) -- (0.39,1.961);
				\node at (-0.507,2.55) {$i_1$};
				\draw[ultra thick] (1.961,0.39) arc (11.25:78.75:2);
				\filldraw (0.39,-1.961) circle (2pt);
				\filldraw (-0.39,-1.961) circle (2pt);
				\draw [dotted,thick] (1.961,0.39) -- (1.961,-0.39);
				\node at (2.35,0.507) {$i_2$};
				\draw[ultra thick] (0.39,-1.961) arc (-78.75:-11.25:2);
				\filldraw (-1.961,-0.39) circle (2pt);
				\filldraw (-1.961,0.39) circle (2pt);
				\draw [dotted,thick] (0.39,-1.961) -- (-0.39,-1.961);
				\node at (0.507,-2.55) {$i_3$};
				\draw[ultra thick] (-1.961,-0.39) arc (-168.75:-101.25:2);
				\filldraw (-0.39,1.961) circle (2pt);
				\filldraw (0.39,1.961) circle (2pt);
				\draw [dotted,thick] (-1.961,-0.39) -- (-1.961,0.39);
				\node at (-2.35,-0.507) {$i_4$};
				\draw [thick] (-0.39,1.961) -- (1.961,0.39);
				\draw [thick] (0.39,1.961) -- (0.39,-1.961);
				\draw [thick] (1.961,-0.39) -- (-1.961,-0.39);
				\draw [thick] (-0.39,-1.961) -- (-1.961,0.39);

				\draw[ultra thick] (5.609,1.961) arc (101.25:168.75:2);
				\filldraw (7.961,0.39) circle (2pt);
				\filldraw (7.961,-0.39) circle (2pt);
				\draw [dotted,thick] (5.609,1.961) -- (6.39,1.961);
				\node at (5.492,2.55) {$i_1$};
				\draw[ultra thick] (7.961,0.39) arc (11.25:78.75:2);
				\filldraw (6.39,-1.961) circle (2pt);
				\filldraw (5.609,-1.961) circle (2pt);
				\draw [dotted,thick] (7.961,0.39) -- (7.961,-0.39);
				\node at (8.35,0.507) {$i_2$};
				\draw[ultra thick] (6.39,-1.961) arc (-78.75:-11.25:2);
				\filldraw (4.038,-0.39) circle (2pt);
				\filldraw (4.038,0.39) circle (2pt);
				\draw [dotted,thick] (6.39,-1.961) -- (5.609,-1.961);
				\node at (6.507,-2.55) {$i_3$};
				\draw[ultra thick] (4.038,-0.39) arc (-168.75:-101.25:2);
				\filldraw (5.609,1.961) circle (2pt);
				\filldraw (6.39,1.961) circle (2pt);
				\draw [dotted,thick] (4.038,-0.39) -- (4.038,0.39);
				\node at (3.649,-0.507) {$i_4$};
				\draw [thick] (5.609,1.961) -- (7.961,0.39);
				\draw [thick] (6.39,1.961) -- (7.961,-0.39);
				\draw [thick] (6.39,-1.961) -- (4.038,-0.39);
				\draw [thick] (5.609,-1.961) -- (4.038,0.39);
				
				\draw[ultra thick] (11.609,1.961) arc (101.25:168.75:2);
				\filldraw (13.961,0.39) circle (2pt);
				\filldraw (13.961,-0.39) circle (2pt);
				\draw [dotted,thick] (11.609,1.961) -- (12.39,1.961);
				\node at (11.492,2.55) {$i_1$};
				\draw[ultra thick] (13.961,0.39) arc (11.25:78.75:2);
				\filldraw (12.39,-1.961) circle (2pt);
				\filldraw (11.609,-1.961) circle (2pt);
				\draw [dotted,thick] (13.961,0.39) -- (13.961,-0.39);
				\node at (14.35,0.507) {$i_2$};
				\draw[ultra thick] (12.39,-1.961) arc (-78.75:-11.25:2);
				\filldraw (10.038,-0.39) circle (2pt);
				\filldraw (10.038,0.39) circle (2pt);
				\draw [dotted,thick] (12.39,-1.961) -- (11.609,-1.961);
				\node at (12.507,-2.55) {$i_3$};
				\draw[ultra thick] (10.038,-0.39) arc (-168.75:-101.25:2);
				\filldraw (11.609,1.961) circle (2pt);
				\filldraw (12.39,1.961) circle (2pt);
				\draw [dotted,thick] (10.038,-0.39) -- (10.038,0.39);
				\node at (9.649,-0.507) {$i_4$};
				\draw [thick] (11.609,1.961) -- (13.961,0.39);
				\draw [thick] (12.39,1.961) -- (10.038,-0.39);
				\draw [thick] (11.609,-1.961) -- (13.961,-0.39);
				\draw [thick] (12.39,-1.961) -- (10.038,0.39);

				\draw[ultra thick] (17.609,1.961) arc (101.25:168.75:2);
				\filldraw (19.961,0.39) circle (2pt);
				\filldraw (19.961,-0.39) circle (2pt);
				\draw [dotted,thick] (17.609,1.961) -- (18.39,1.961);
				\node at (17.492,2.55) {$i_1$};
				\draw[ultra thick] (19.961,0.39) arc (11.25:78.75:2);
				\filldraw (18.39,-1.961) circle (2pt);
				\filldraw (17.609,-1.961) circle (2pt);
				\draw [dotted,thick] (19.961,0.39) -- (19.961,-0.39);
				\node at (20.35,0.507) {$i_2$};
				\draw[ultra thick] (18.39,-1.961) arc (-78.75:-11.25:2);
				\filldraw (16.038,-0.39) circle (2pt);
				\filldraw (16.038,0.39) circle (2pt);
				\draw [dotted,thick] (18.39,-1.961) -- (17.609,-1.961);
				\node at (18.507,-2.55) {$i_3$};
				\draw[ultra thick] (16.038,-0.39) arc (-168.75:-101.25:2);
				\filldraw (17.609,1.961) circle (2pt);
				\filldraw (18.39,1.961) circle (2pt);
				\draw [dotted,thick] (16.038,-0.39) -- (16.038,0.39);
				\node at (15.649,-0.507) {$i_4$};
				\draw [thick] (17.609,1.961) -- (19.961,0.39);
				\draw [thick] (18.39,1.961) -- (17.609,-1.961);
				\draw [thick] (16.038,-0.39) -- (18.39,-1.961);
				\draw [thick] (19.961,-0.39) -- (16.038,0.39);
				
				\draw[ultra thick] (2.609,-5.038) arc (101.25:168.75:2);
				\filldraw (4.961,-6.609) circle (2pt);
				\filldraw (4.961,-7.39) circle (2pt);
				\draw [dotted,thick] (2.609,-5.038) -- (3.39,-5.038);
				\node at (2.492,-4.449) {$i_1$};
				\draw[ultra thick] (4.961,-6.609) arc (11.25:78.75:2);
				\filldraw (3.39,-8.961) circle (2pt);
				\filldraw (2.609,-8.961) circle (2pt);
				\draw [dotted,thick] (4.961,-6.609) -- (4.961,-7.39);
				\node at (5.35,-6.492) {$i_2$};
				\draw[ultra thick] (3.39,-8.961) arc (-78.75:-11.25:2);
				\filldraw (1.038,-7.39) circle (2pt);
				\filldraw (1.038,-6.609) circle (2pt);
				\draw [dotted,thick] (3.39,-8.961) -- (2.609,-8.961);
				\node at (3.507,-9.55) {$i_3$};
				\draw[ultra thick] (1.038,-7.39) arc (-168.75:-101.25:2);
				\filldraw (2.609,-5.038) circle (2pt);
				\filldraw (3.39,-5.038) circle (2pt);
				\draw [dotted,thick] (1.038,-7.39) -- (1.038,-6.609);
				\node at (0.649,-7.507) {$i_4$};
				\draw [thick] (2.609,-5.038) -- (4.961,-6.609);
				\draw [thick] (3.39,-5.038) -- (2.609,-8.961);
				\draw [thick] (1.038,-7.39) -- (4.961,-7.39);
				\draw [thick] (3.39,-8.961) -- (1.038,-6.609);

				\draw[ultra thick] (8.609,-5.038) arc (101.25:168.75:2);
				\filldraw (10.961,-6.609) circle (2pt);
				\filldraw (10.961,-7.39) circle (2pt);
				\draw [dotted,thick] (8.609,-5.038) -- (9.39,-5.038);
				\node at (8.492,-4.449) {$i_1$};
				\draw[ultra thick] (10.961,-6.609) arc (11.25:78.75:2);
				\filldraw (9.39,-8.961) circle (2pt);
				\filldraw (8.609,-8.961) circle (2pt);
				\draw [dotted,thick] (10.961,-6.609) -- (10.961,-7.39);
				\node at (11.35,-6.492) {$i_2$};
				\draw[ultra thick] (9.39,-8.961) arc (-78.75:-11.25:2);
				\filldraw (7.038,-7.39) circle (2pt);
				\filldraw (7.038,-6.609) circle (2pt);
				\draw [dotted,thick] (9.39,-8.961) -- (8.609,-8.961);
				\node at (9.492,-9.55) {$i_3$};
				\draw[ultra thick] (7.038,-7.39) arc (-168.75:-101.25:2);
				\filldraw (8.609,-5.038) circle (2pt);
				\filldraw (9.39,-5.038) circle (2pt);
				\draw [dotted,thick] (7.038,-7.39) -- (7.038,-6.609);
				\node at (6.649,-7.507) {$i_4$};
				\draw [thick] (8.609,-5.038) -- (9.39,-8.961);
				\draw [thick] (10.961,-7.39) -- (7.038,-7.39);
				\draw [thick] (8.609,-8.961) -- (9.39,-5.038);
				\draw [thick] (10.961,-6.609) -- (7.038,-6.609);
				
				\draw[ultra thick] (14.609,-5.038) arc (101.25:168.75:2);
				\filldraw (16.961,-6.609) circle (2pt);
				\filldraw (16.961,-7.39) circle (2pt);
				\draw [dotted,thick] (14.609,-5.038) -- (15.39,-5.038);
				\node at (14.492,-4.449) {$i_1$};
				\draw[ultra thick] (16.961,-6.609) arc (11.25:78.75:2);
				\filldraw (15.39,-8.961) circle (2pt);
				\filldraw (14.609,-8.961) circle (2pt);
				\draw [dotted,thick] (16.961,-6.609) -- (16.961,-7.39);
				\node at (17.35,-6.492) {$i_2$};
				\draw[ultra thick] (15.39,-8.961) arc (-78.75:-11.25:2);
				\filldraw (13.038,-7.39) circle (2pt);
				\filldraw (13.038,-6.609) circle (2pt);
				\draw [dotted,thick] (15.39,-8.961) -- (14.609,-8.961);
				\node at (15.507,-9.55) {$i_3$};
				\draw[ultra thick] (13.038,-7.39) arc (-168.75:-101.25:2);
				\filldraw (14.609,-5.038) circle (2pt);
				\filldraw (15.39,-5.038) circle (2pt);
				\draw [dotted,thick] (13.038,-7.39) -- (13.038,-6.609);
				\node at (12.649,-7.507) {$i_4$};
				\draw [thick] (14.609,-5.038) -- (14.609,-8.961);
				\draw [thick] (13.038,-7.39) -- (16.961,-7.39);
				\draw [thick] (15.39,-8.961) -- (15.39,-5.038);
				\draw [thick] (16.961,-6.609) -- (13.038,-6.609);
				
				{\normalsize
					\node at (0,-3.5) {$\rq{1}$};
					\node at (6,-3.5) {$\rq{2}$};
					\node at (12,-3.5) {$\rq{3}$};
					\node at (18,-3.5) {$\rq{4}$};
					\node at (3,-10.5) {$\rq{5}$};
					\node at (9,-10.5) {$\rq{10}$};
					\node at (15,-10.5) {$\rq{25}$};
				}
				
			\end{tikzpicture}
		}
		\caption{Orbits of \opt{4}}
		\label{fig:orbits4}
	\end{center}
\end{figure}

We can list all the orbits by specifying just one element, called the {\em representative}, for each of them, since the other elements can be obtained by the action of ${\cal G}$
on the representative. By convention, we have  chosen as the representative the smallest (in lexicographic order) scheme of the orbit.
In Fig. \ref{fig:orbits4} we illustrate the representatives of the 7 orbits.

\newcommand{\rtrei}{r_1}
\newcommand{\rtreii}{r_2}
\newcommand{\rtreiii}{r_3}
\newcommand{\rtreiv}{r_4}


 \newcommand{\STOP}{{\tt STOP}}
 \newcommand{\FALSE}{{\tt false}}
 \newcommand{\BEST}{{\tt BESTIMPR}}
\newcommand{\comment}[1]{\ \ /{\tt *} \textrm{#1}\ {\tt *}/}
\newcommand{\FOR}{{\bf for\ }}
\newcommand{\DO}{{\bf do\ }}
\newcommand{\TO}{{\bf to\ }}
\newcommand{\IF}{{\bf if\ }}
\newcommand{\THEN}{{\bf then\ }}
\newcommand{\ELSE}{{\bf else\ }}
\newcommand{\ENDIF}{{\bf endif\ }}
\newcommand{\ENDFOR}{{\bf endfor\ }}
\newcommand{\WHILE}{{\bf while\ }}
\newcommand{\ENDWHILE}{{\bf endwhile\ }}
\newcommand{\RETURN}{{\bf return\ }}

\newcommand{\INPUT}[1]{\vskip 0.11cm\hskip 2.5cm {\bf Input:} #1}
\newcommand{\LOCAL}[1]{\vskip 0.11cm\hskip 2.5cm {\bf Local:} #1}
\newcommand{\OUTPUT}[1]{\vskip 0.11cm\hskip 2.5cm {\bf Output:} #1}

\newcommand{\fst}[1]{{\tt fst}(#1)}
\newcommand{\scn}[1]{{\tt snd}(#1)}
\newcommand{\sel}[1]{{\tt selection}(#1)}
\newcommand{\minval}[1]{{\tt minval}(#1)}
\newcommand{\maxval}[1]{{\tt maxval}(#1)}
\newcommand{\rangeAB}[1]{{\tt range1st2nd}(#1)}
\newcommand{\rangeC}[1]{{\tt range3rd}(#1)}
\newcommand{\HSIZE}{H.\texttt{SIZE}}

\section{Methods for \opt{4} exploration in the literature}
\label{sec:literature}

In this section we describe the cubic de Berg et al. \cite{woeginger}  algorithm and the quadratic approach by Glover \cite{Glover1996}.
It must be remarked that Glover's algorithm has a very restricted scope, i.e., it 
can be applied only to 3 out of 25 reinsertion schemes, while de Berg et al. procedure is valid for
all \opt{4} moves.

\newcommand{\costd}{\textrm{cost}_d}
\newcommand{\costc}{\textrm{cost}_c}
\newcommand{\bestA}{\textrm{bestA}}
\newcommand{\bestB}{\textrm{bestB}}
\newcommand{\bestV}{\textrm{bestV}}
\renewcommand{\myquote}[1]{``#1''}

\subsection{de Berg et al. $\Theta(n^3)$ dynamic programming approach}

\begin{figure}[h]
	\begin{center}
		{\footnotesize
			\begin{tikzpicture}[scale=0.6]
				
				\draw[ultra thick] (11.609,1.961) arc (101.25:168.75:2);
				\filldraw (13.961,0.39) circle (2pt);
				\filldraw (13.961,-0.39) circle (2pt);
				\draw [dotted,thick] (11.609,1.961) -- (12.39,1.961);
				\node at (11.492,2.55) {$a_1$};
				\draw[ultra thick] (13.961,0.39) arc (11.25:78.75:2);
				\filldraw (12.39,-1.961) circle (2pt);
				\filldraw (11.609,-1.961) circle (2pt);
				\draw [dotted,thick] (13.961,0.39) -- (13.961,-0.39);
				\node at (14.35,0.507) {$b_1$};
				\node at (12.75,0.85) {$l_1$};
				\draw[ultra thick] (12.39,-1.961) arc (-78.75:-11.25:2);
				\filldraw (10.038,-0.39) circle (2pt);
				\filldraw (10.038,0.39) circle (2pt);
				\draw [dotted,thick] (12.39,-1.961) -- (11.609,-1.961);
				\node at (12.507,-2.55) {$a_2$};
				\node at (12.75,-0.75) {$l_2$};
				\node at (10.9,-0.9) {$l_3$};
				\node at (11.0,1.1) {$l_4$};
				\draw[ultra thick] (10.038,-0.39) arc (-168.75:-101.25:2);
				\filldraw (11.609,1.961) circle (2pt);
				\filldraw (12.39,1.961) circle (2pt);
				\draw [dotted,thick] (10.038,-0.39) -- (10.038,0.39);
				\node at (9.649,-0.507) {$b_2$};
				\draw [thick] (11.609,1.961) -- (13.961,0.39);
				\draw [thick] (12.39,1.961) -- (10.038,-0.39);
				\draw [thick] (11.609,-1.961) -- (13.961,-0.39);
				\draw [thick] (12.39,-1.961) -- (10.038,0.39);

				{\normalsize
					\node at (12,-3.5) {$\rq{3}$};
				}
				
			\end{tikzpicture}
		}
		\caption{An example reinsertion scheme for de Berg's et al. algorithm}
		\label{fig:berg}
	\end{center}
\end{figure}
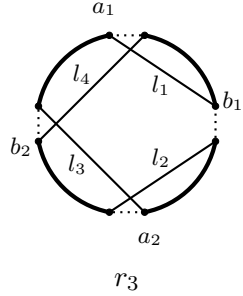

For a given reinsertion scheme and selection  $(i_1,i_2,i_3,i_4)$, let  $L$ be the set of edges inserted by the scheme and  $e_k=\{i_k,i_k\piu 1\}$, for $k=1,\ldots,4$, be the edges removed.
We say that two nodes $i_s$ and $i_t$ are {\em independent} if no edge $l\in L$ is incident on both $e_s$ and $e_t$.
For instance, in Fig. \ref{fig:berg}, the nodes $a_1$ and $a_2$ are independent, while $a_1$ and $b_1$ are not, etc. 
It can be shown that, for every reinsertion 
scheme, the selection nodes
can be partitioned into two groups say $A=\{a_1, a_2\}$ and $B=\{b_1, b_2\}$
of independent nodes, and the edges $L$ can
be partitioned into $\{l_1,l_2\} \cup \{l_3,l_4\}$ in such a way that the cost of the move is 
\[
\begin{matrix*}[l]
	c(a_1,a_1\piu 1) + c(a_2,a_2\piu 1) + 
	[c(b_1,b_1\piu 1) - c(l_1) - c(l_2)] + 
	[c(b_2,b_2\piu 1) - c(l_3) - c(l_4)].
\end{matrix*}
\]
Given a scheme $r$, to find the best move we can consider all possibilities for the choice of $a_1$ and $a_2$ (there are $\Theta(n^2)$ choices). For each such choice, we can then try to find the best possible completion by optimally placing the remaining two nodes $b_1$ and $b_2$ of the selection.
This can be done via a dynamic programming procedure in time $\Theta(n)$ rather than $\Theta(n^2)$ so that the overall procedure for finding the best move for $r$ has complexity $\Theta(n^3)$.

Given the placement of the two fixed nodes $a_1$ and $a_2$, the value of a selection is split into three parts, i.e., a constant part $c(a_1,a_1\piu 1)+c(a_2,a_2\piu 1)$ depending only on $a_1,a_2$ plus two contributions 
\[
\tilde c_1(b_1|a_1,a_2) := c(b_1,b_1\piu 1) - c(l_1) - c(l_2)
\]
and 
\[
\tilde c_2(b_2|a_1,a_2) := c(b_2,b_2\piu 1) - c(l_3) - c(l_4)
\]
which we seek to maximize. Each of these contribution accounts for the cost of one removed edge minus the cost of two inserted edges. Since the positions $b_1$ and $b_2$ are independent,  all the inserted edges get counted exactly once.
For instance, with respect to the reinsertion scheme $r_3$ in Fig. \ref{fig:berg}, in which  $a_1=i_1$ and $a_2=i_3$,  it is
\[
\tilde c_1(b_1|a_1,a_2) = c(b_1,b_1\piu 1) - c(b_1,a_1) - c(b_1\piu 1,a_2\piu 1) 
\]
and 
\[
{\tilde c}_2(b_2|a_1,a_2) = c(b_2,b_2\piu 1) - c(b_2,a_2\piu 1) - c(b_2\piu 1,a_2).
\]
\ \\
In order to find the placement of $b_1$ and $b_2$ which maximizes $\tilde c_1(b_1|a_1,a_2)+\tilde c_1(b_2|a_1,a_2)$, we consider a general situation in which there is a range of consecutive positions $\{\min_1,\ldots, \max_1\}$ where we have to place $b_1$ and a range $\{\min_2,\ldots, \max_2\}$ for $b_2$. These ranges depend on the scheme $r$. E.g., in our example, it is $\min_1= a_1+2$, $\max_1=a_2-2$, and $\min_2 = a_2 + 2$,  $\max_2=\bar{n} - \PP{a_1=0}$. 

For $j\in \{\min_1,\ldots,\max_1\}$ let us call $V_1[j]$ the best value of a feasible partial selection $\{a_1,a_2,b_1\}$ for which $b_1\in \{\min_1, \ldots, j\}$, and let $\bestV_1[j]$ be the corresponding choice for $b_1$. 
We have the general recurrence, for $\min_1 < j \le \max_1$,
\[
V_1[j] = \max\{V_1[j-1],\, {\tilde c}_1(j|a_1,a_2)\}
\]
where $\bestV_1[j]\leftarrow \bestV_1[j-1]$ if the
max is achieved by first term, and  $\bestV_1[j]\leftarrow  j$ otherwise.

Similarly, let $V_2[j]$ the best value of a feasible selection $\{a_1,a_2,b_1,b_2\}$ for which $b_2\in \{\min_2,\ldots, j\}$ and  let $\bestV_1[j]$ be the best value for $b_2$.  We have the general recurrence, for $\min_2 < j \le \max_2$,
\[
V_2[j] = \max\{V_2[j-1],\, {\tilde c}_2(j|a_1,a_2) + V_1[\min \{ j-2, \textrm{max}_1\}]\}.
\]
where $\bestV_2[j]\leftarrow \bestV_2[j-1]$ if the
max is achieved by first term, and  $\bestV_2[j]\leftarrow  j$ otherwise.

The optimal solution value for $a_1$, $a_2$ is found in $V_2[\max_2]$ and can be computed in time $\Theta(n)$ by filling the arrays $V_1[\cdot]$ and $V_2[\cdot]$. The optimal completion of $a_1, a_2$ is
$b_2:= \bestV_2[\textrm{max}_2]$ and $b_1:=\bestV_1[\min \{ b_2-2, \textrm{max}_1\}]$.

\subsection{Glover's $\Theta(n^2)$ algorithm}

\begin{figure}[h]
	\begin{center}
		{\footnotesize
			\begin{tikzpicture}[scale=0.6]
				
				\draw[ultra thick] (8.609,-5.038) arc (101.25:168.75:2);
				\filldraw (10.961,-6.609) circle (2pt);
				\filldraw (10.961,-7.39) circle (2pt);
				\draw [dotted,thick] (8.609,-5.038) -- (9.39,-5.038);
				\node at (8.492,-4.449) {$a$};
				\node at (9.507,-4.449) {$a + 1$};
				\draw[ultra thick] (10.961,-6.609) arc (11.25:78.75:2);
				\filldraw (9.39,-8.961) circle (2pt);
				\filldraw (8.609,-8.961) circle (2pt);
				\draw [dotted,thick] (10.961,-6.609) -- (10.961,-7.39);
				\draw[ultra thick] (9.39,-8.961) arc (-78.75:-11.25:2);
				\filldraw (7.038,-7.39) circle (2pt);
				\filldraw (7.038,-6.609) circle (2pt);
				\draw [dotted,thick] (9.39,-8.961) -- (8.609,-8.961);
				\node at (9.492,-9.55) {$b$};
				\node at (8.492,-9.55) {$b + 1$};
				\draw[ultra thick] (7.038,-7.39) arc (-168.75:-101.25:2);
				\filldraw (8.609,-5.038) circle (2pt);
				\filldraw (9.39,-5.038) circle (2pt);
				\draw [dotted,thick] (7.038,-7.39) -- (7.038,-6.609);
				\draw [thick] (8.609,-5.038) -- (9.39,-8.961);
				\draw [thick] (10.961,-7.39) -- (7.038,-7.39);
				\draw [thick] (8.609,-8.961) -- (9.39,-5.038);
				\draw [thick] (10.961,-6.609) -- (7.038,-6.609);
				
				\draw[ultra thick] (14.609,-5.038) arc (101.25:168.75:2);
				\filldraw (16.961,-6.609) circle (2pt);
				\filldraw (16.961,-7.39) circle (2pt);
				\draw [dotted,thick] (14.609,-5.038) -- (15.39,-5.038);
				\node at (14.492,-4.449) {$a$};
				\node at (15.507,-4.449) {$a+ 1$};
				\draw[ultra thick] (16.961,-6.609) arc (11.25:78.75:2);
				\filldraw (15.39,-8.961) circle (2pt);
				\filldraw (14.609,-8.961) circle (2pt);
				\draw [dotted,thick] (16.961,-6.609) -- (16.961,-7.39);
				\draw[ultra thick] (15.39,-8.961) arc (-78.75:-11.25:2);
				\filldraw (13.038,-7.39) circle (2pt);
				\filldraw (13.038,-6.609) circle (2pt);
				\draw [dotted,thick] (15.39,-8.961) -- (14.609,-8.961);
				\node at (15.507,-9.55) {$b$};
				\node at (14.492,-9.55) {$b+ 1$};
				\draw[ultra thick] (13.038,-7.39) arc (-168.75:-101.25:2);
				\filldraw (14.609,-5.038) circle (2pt);
				\filldraw (15.39,-5.038) circle (2pt);
				\draw [dotted,thick] (13.038,-7.39) -- (13.038,-6.609);
				\draw [thick] (14.609,-5.038) -- (14.609,-8.961);
				\draw [thick] (13.038,-7.39) -- (16.961,-7.39);
				\draw [thick] (15.39,-8.961) -- (15.39,-5.038);
				\draw [thick] (16.961,-6.609) -- (13.038,-6.609);
				
				{\normalsize
					\node at (9,-10.5) {${\cal O}_{6}$};
					\node at (15,-10.5) {${\cal O}_{7}$};
				}
				
			\end{tikzpicture}
		}
		\caption{Orbits for Glover's algorithm}
		\label{fig:glover}
	\end{center}
\end{figure}
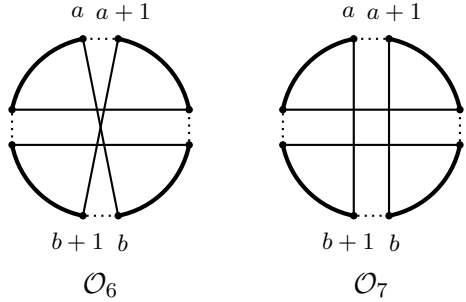

In this section we briefly describe Glover's $\Theta(n^2)$ algorithm
for finding the best \opt{4} move. The algorithm works only for a restricted set of reinsertion
schemes, i.e., $r_{10}$, $r_{16}$ (orbit ${\cal O}_{6}$) and $r_{25}$ (orbit ${\cal O}_7$). For convenience, we report these orbits in Fig. \ref{fig:glover}. 

We use a different (simpler but equivalent) explanation than in Glover's paper, based
on dynamic programming. The description is recursive, but in the
implementation, by using memoization, the recursion is avoided. The
running time is $\Theta(n^2)$.

\paragraph{Orbit ${\cal O}_7$.}

Orbit ${\cal O}_7 = \{r_{25}\}$ (also called the \myquote{two bridges} reinsertion scheme, because of 
its peculiar shape)  can be seen as the combination of
two independent moves of two edges each.
Each such move, let us call it $M(a,b)$, removes the edges $\edg{a}{a+1}$ and $\edg{b}{b+1}$ and
introduces $\edg{a}{b+1}$ and $\edg{a+1}{b}$, thus disconnecting the tour. Let us call the cost of this disconnecting move
\[
\costd(a,b) := c(a,a+1) + c(b,b+1) - (c(a,b+1) + c(a+1,b)).
\]

We define two dynamic programming tables $A[i,j]$ and $B[i,j]$, for $0\le i < j \le \bar n$. The meaning of $A$ is
$A[i,j]$=\myquote{value of the best possible move $M(a,j)$ with $a\le i$}. I.e., the move removes
 an edge $\edg{a}{a+1}$ in the interval $0,\ldots,i$ together with $\edg{j}{j+1}$. The best value for $a$ is saved in an array $\bestA[i,j]$. We have the following recursion:
\[
A[i,j] = \max\{ \costd(i,j), A[i-1,j] \}
\]
where the first case is when we remove $\edg{i}{i+1}$, while the
second case is when we do not. Base case occurs when $i=0$.

The meaning of $B$ is $B[i,j]$=\myquote{value of the best possible move $M(a,b)$ with $0\le a\le i-2$ and $i+2 \le b\le j-2$}. I.e., the move removes an edge $\edg{a}{a+1}$ to the left of $i$ (in the interval $0,\ldots,i-2$) together with an edge $\edg{b}{b+1}$ to the right of $i$ but to the left of $j$ (i.e., in the interval $i+2,\ldots,j-2$). The best value for $(a,b)$ is saved in $\bestB[i,j]$.
We have the following recursion:
\[
B[i,j] = \max\{ A[i-2,j-2], B[i,j-1] \}
\]
where the first case is when we do remove the rightmost possible arc, i.e., $\edg{j-2}{j-1}$, while the
second case is when we do not. Base case occurs when $j-2=i+2$.
When the max is the first term, $\bestB[i,j] := (\bestA[i-2,j-2], j-2)$. When
it is the second, $\bestB[i,j]:=\bestB[i,j-1]$.

At this point, we can state the whole dynamic programming algorithm,
by guessing two of the edges of the \opt{4} move, namely the
2nd $\edg{i_2}{i_2+1}$ and the 4th $\edg{i_4}{i_4+1}$ and completing in the best possible way with the 1st and 3rd. We compute $B[x,y]$ in time $O(n^2)$ and then, still in time $O(n^2)$
\[
OPT(r_{25}) := \max \{\costd(i_2,i_4) + B[i_2,i_4] : 2\le i_2 < i_4 -4 \le i_4 < \bar n \}.
\]

If $i'_2$ and $i'_4$ realize the above maximum, then the
best move is completed by setting
\[
(i'_1, i'_3) := \bestB[i'_2, i'_4].
\]

\paragraph{Orbit ${\cal O}_6$.}

Orbit ${\cal O}_6$ is made by one \myquote{crossed} bridge and one \myquote{parallel} bridge, intertwined. If the first bridge (the one indexed by $i_1$ and $i_3$) is the crossed one, we have reinsertion scheme $r_{10}$, otherwise we have $r_{16}$. The cost of 
a crossed bridge indexed by $a$ and $b$ is
\[
\costc(a,b) := c(a,a+1) + c(b,b+1) - (c(a,b) + c(a+1,b+1)).
\]

We can use a similar dynamic program as before. In the previous section, $B[x,y]$ finds the best placement of a \myquote{parallel} bridge
given $x$ and $y$, and is based on the cost $\costd$. If
we were to replace $\costd()$ with $\costc()$ in the definition of $A[,]$, then  $B[x,y]$ would find the best placement of a \myquote{crossed}
 bridge given $x$ and $y$. So let us assume that $B^c$ is the version
of $B$ using $\costc()$, while 
$B^d$ is the version
of $B$ using $\costd()$. Then
\[
OPT(r_{16}) := \max \{\costc(i_2,i_4) + B^d[i_2,i_4] : 2\le i_2 < i_4 -4 \le i_4 < \bar n \}
\]
and 
\[
OPT(r_{10}) := \max \{\costd(i_2,i_4) + B^c[i_2,i_4] : 2\le i_2 < i_4 -4 \le i_4 < \bar n \}.
\]

\bibliographystyle{plain}
\bibliography{tspbib}

\begin{thebibliography}{1}

\bibitem{AaLe97}
Emile Aarts and Jan~K. Lenstra, editors.
\newblock {\em Local Search in Combinatorial Optimization}.
\newblock John Wiley \& Sons, Inc., New York, NY, USA, 1st edition, 1997.

\bibitem{Croes2opt}
G.~A. Croes.
\newblock A method for solving traveling-salesman problems.
\newblock {\em Operations Research}, 6(6):791--812, 1958.

\bibitem{woeginger}
Mark de~Berg, Kevin Buchin, Bart M.~P. Jansen, and Gerhard~J. Woeginger.
\newblock Fine-grained complexity analysis of two classic {TSP} variants.
\newblock In {\em 43rd International Colloquium on Automata, Languages, and
  Programming, {ICALP} 2016, July 11-15, 2016, Rome, Italy}, pages 5:1--5:14,
  2016.

\bibitem{Glover1996}
Fred Glover.
\newblock Finding a best traveling salesman 4-{OPT} move in the same time as a
  best 2-{OPT} move.
\newblock {\em Journal of Heuristics}, 2(2):169--179, Sep 1996.

\bibitem{PapSte}
Christos H.~Papadimitriou and Kenneth Steiglitz.
\newblock {\em Combinatorial Optimization: Algorithms and Complexity}.
\newblock Prentice Hall, 01 1982.

\bibitem{SW}
Steiglitz Kenneth and Weiner Peter.
\newblock Some improved algorithms for computer solution of the traveling
  salesman problem.
\newblock In {\em Proc. 6th annual Allerton Conf. on System and System Theory},
  pages 814--821. University of Illinois, Urbana, 1968.

\bibitem{LD20}
Giuseppe Lancia and Marcello Dalpasso.
\newblock Finding the best 3-{OPT} move in subcubic time.
\newblock {\em Algorithms}, 13(11):306--332, 2020.

\bibitem{Lin65}
S.~Lin.
\newblock Computer solutions of the traveling salesman problem.
\newblock {\em The Bell System Technical Journal}, 44(10):2245--2269, Dec 1965.

\end{thebibliography}

\end{document}